\documentclass[preprint2]{aastex}
\usepackage{graphicx}
\newcommand{\etal}{{et al.}}
\newcommand{\ddeg}{$^\circ$}
\newcommand{\figwidth}{75mm}
\newcommand{\be}{\begin{equation}}
\newcommand{\ee}{\end{equation}}
\begin{document}
   \title{First whole atmosphere night-time seeing measurements at Dome C, Antarctica}

   \author{A. Agabi, E. Aristidi, M. Azouit, E. Fossat, 
   F. Martin, 
 T. Sadibekova,
J. Vernin, 
   A. Ziad}

   \affil{Laboratoire Universitaire d'Astrophysique de Nice,
              Universit\'e de Nice Sophia Antipolis, Parc Valrose,
	      F-06108 Nice Cedex 2}

   \begin{abstract}
   We report site testing results obtained in night-time during the polar autumn and winter at Dome C. These results were collected during the first Concordia winterover by A. Agabi. They are based upon seeing and isoplanatic angle monitoring, as well as in-situ balloon measurements of the refractive index structure constant profiles $C_n^2(h)$. Atmosphere is  divided into two regions: (i) a 36~m high surface layer responsible of 87\% of the turbulence and (ii) a very stable free atmosphere above with a median seeing of 0.36$\pm$0.19 arcsec at an elevation of $h=30$~m. The median seeing measured with a DIMM placed on top of a 8.5~m high tower is 1.3$\pm$0.8 arcsec.

\end{abstract} 



\section{Introduction}

The French and Italian polar station Concordia based at Dome C on the Antarctic plateau, has just completed its construction. This location (75°S, 123°E) has remarkable properties due to its position on the top of a local maximum of the plateau, at an elevation of 3250~m. Low wind speeds (Aristidi \etal , 2005b) as well as long time periods of clear sky and low sky brightness in the infrared range (Walden \etal , 2005) makes it one of the best candidates for future installation of a large astronomical observatory. Several site testing programs have been undertaken at the end of the 90's to qualify the site. Concordiastro is the program conducted by our group and its goal is to characterize this site for high angular resolution astronomy in the visible. Recent results in summer (Aristidi \etal, 2005a, 2005b) have shown the potential of this site for solar astronomy with a median seeing of 0.54~arcsec in daytime. Another group from the University of New South Wales has installed various instruments in an automated station (AASTINO) and published last year (Lawrence \etal , 2004) some very enthousiastic seeing estimations of 0.27~arcsec, based on stellar scintillation and acoustic radar. 

The first winterover has officially started on Feb. 15, 2005. Among the 13 winterers, A.~Agabi, the project manager of Concordiastro is spending one year at the site to run the experiments, described hereafter in section~\ref{par:matos}. Dome C being at 15\ddeg of the South Pole, there is a transition period of 2$\frac{1}{2}$ months  from summer to winter where the Sun rises and sets. First sunset was observed on Feb.~16 and the Sun totally disappeared from the sky on May~4. During this period it was observed a dramatic drop of the ground temperature (roughly from -30\ddeg C in summer to -70\ddeg C in autumn) and the establishment of an almost permanent inversion layer with a very steep temperature gradient that led to strong optical turbulence near the ground. We present here the temperature and $C_n^2$ profiles obtained from 16 balloon-borne experiments from March 15th to August 1st. Optical parameters (seeing $\epsilon$, isoplanatic angle $\theta_0$ and coherence time $\tau_0$ are derived from these profiles. We show also the results of continuous ground seeing monitoring from March to June 2005, and isoplanatic angle monitoring in May 2005. The winterover is not finished yet, and this paper can be considered as a first look at the data. A more detailed study will be published later when the full season data set will be collected.


\section{Experiments}
\label{par:matos}
Two experiments are currently operated at Dome C to measure the turbulence. The first one aims at monitoring the seeing and the isoplanatic angle. It is based on two small telescopes located at two different heights above the ice. A seeing monitor (DIMM type, Aristidi et al., 2005a), on the top of a platform $h=8.5$~m above ground, provides a seeing value every 2~mn. Another telescope at height $h=3.5$~m monitors either the seeing or the isoplanatic angle. Switching between the two modes is made manually by using different pupil masks. A detailled description is given in Aristidi \etal , 2005a. Monitoring the seeing at two different heights allows to infer the influence of the 5~m thick ground layer (3.5~m$\le h\le$8.5~m). This influence is very faint on the isoplanatic angle, mostly sensitive to high altitude turbulence. In mid-July, a third seeing monitor has been placed onto the roof of the calm building of Concodia, at elevation $h=20$~m. All values (seeing and isoplanatic angle) are computed at wavelength $\lambda=500$~nm and zenithal distance $z=0$\ddeg 

The second experiment consists in {\em in situ} measurements of thermal fluctuations using balloon-borne microthermal sensors. The principle and the performances of these microthermal measurements is detailed in Azouit and Vernin (2005) (see also Marks \etal , 1999). The balloon scans the atmosphere between the ground and an altitude of 15--20~km, sending data every 1--2 seconds. This corresponds to a vertical resolution around 5--10~m, depending on the ascent speed. Each $C_n^2$ value provided by the sensors has an accuracy around 20--25\% . The lower limite of detectable $C_n^2$ is around $10^{-20}$~m$^{-2/3}$. Detailed informations about the microthermal experiment (including balloon wake effects) can be found in Azouit and Vernin (2005).


\section{Results}
\label{par:result}
\subsection{Turbulence profiles from balloon measurements}
\begin{figure*}
\includegraphics[width=\textwidth]{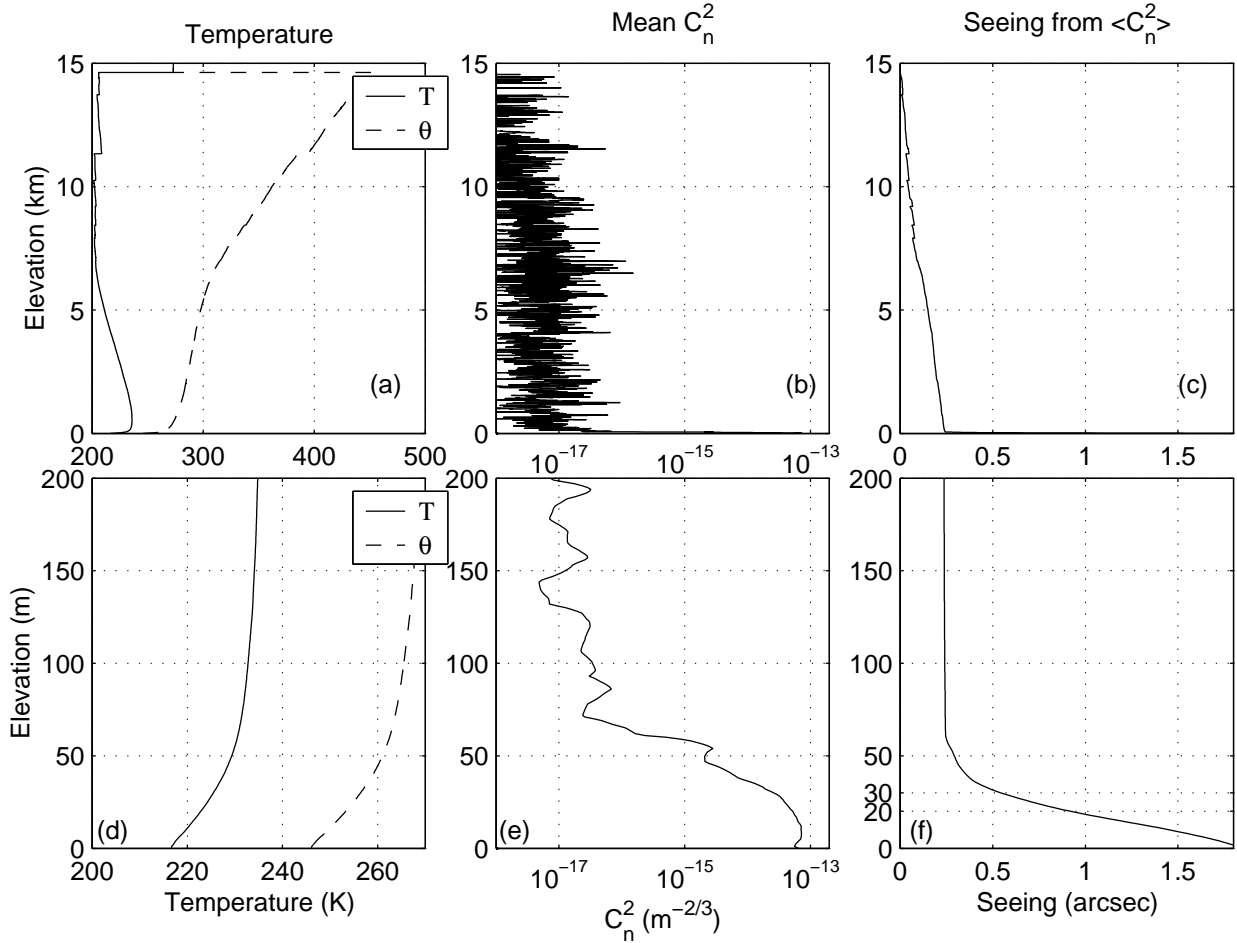}
\caption{Average vertical profiles of (a) the temperature $T$ and potential temperature $\theta$, (b) the refractive index structure constant $C_n^2$. Curves (a) and (b) correspond to averages over 16 balloon data between March 15th and August 1st. (c) is the average seeing  deduced from individual $C_n^2$ profiles. Discontinuities in the temperature plot (a)  are due to the increasingly insufficient number of flights reaching high altitudes. Plots (d), (e) and (f) show the same quantities on the first 200~m over the ground.}
\label{fig:cn2vsalt}
\end{figure*}

\begin{figure*}
\includegraphics[width=\figwidth]{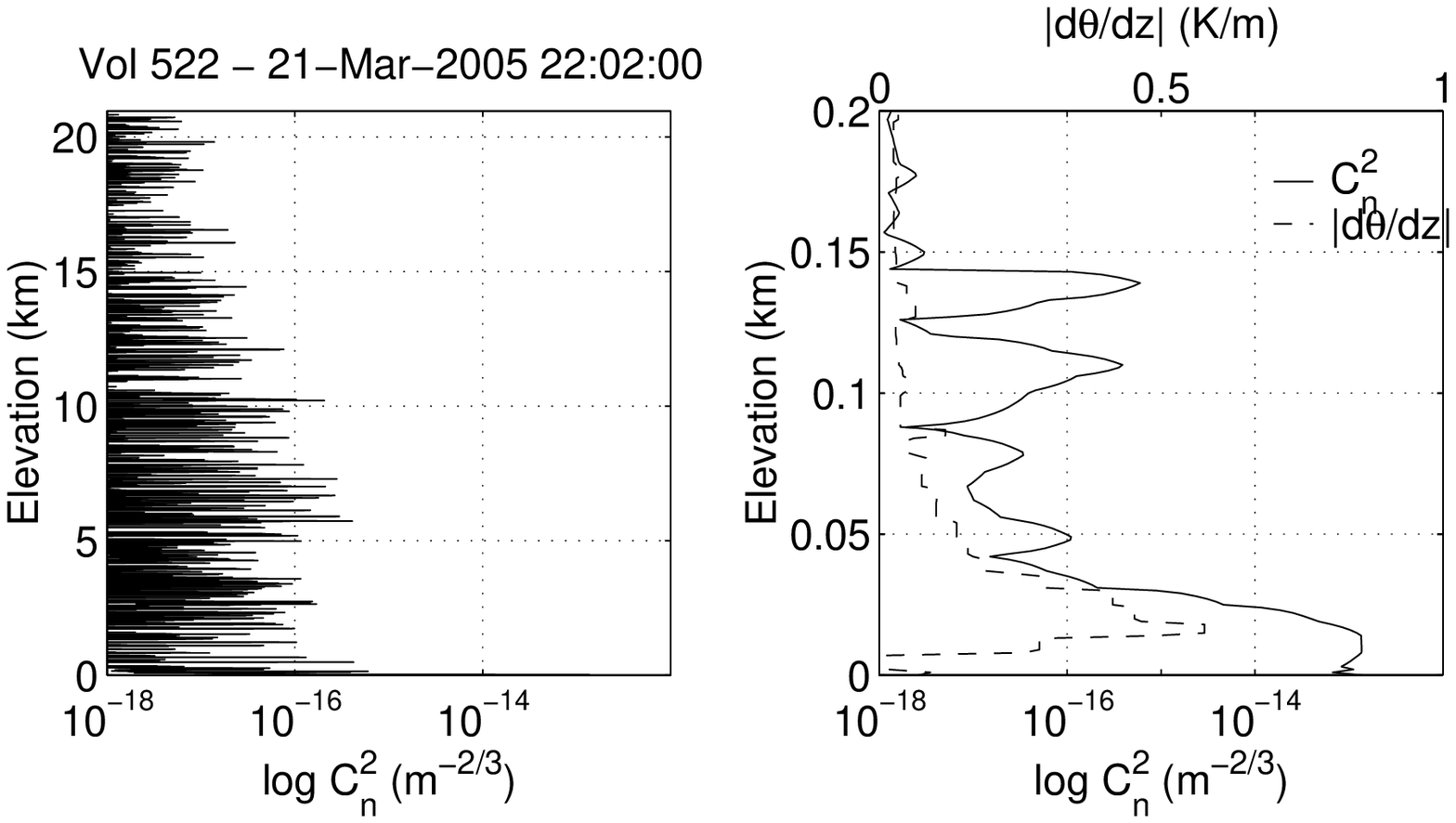}\ \hskip 1cm
\includegraphics[width=\figwidth]{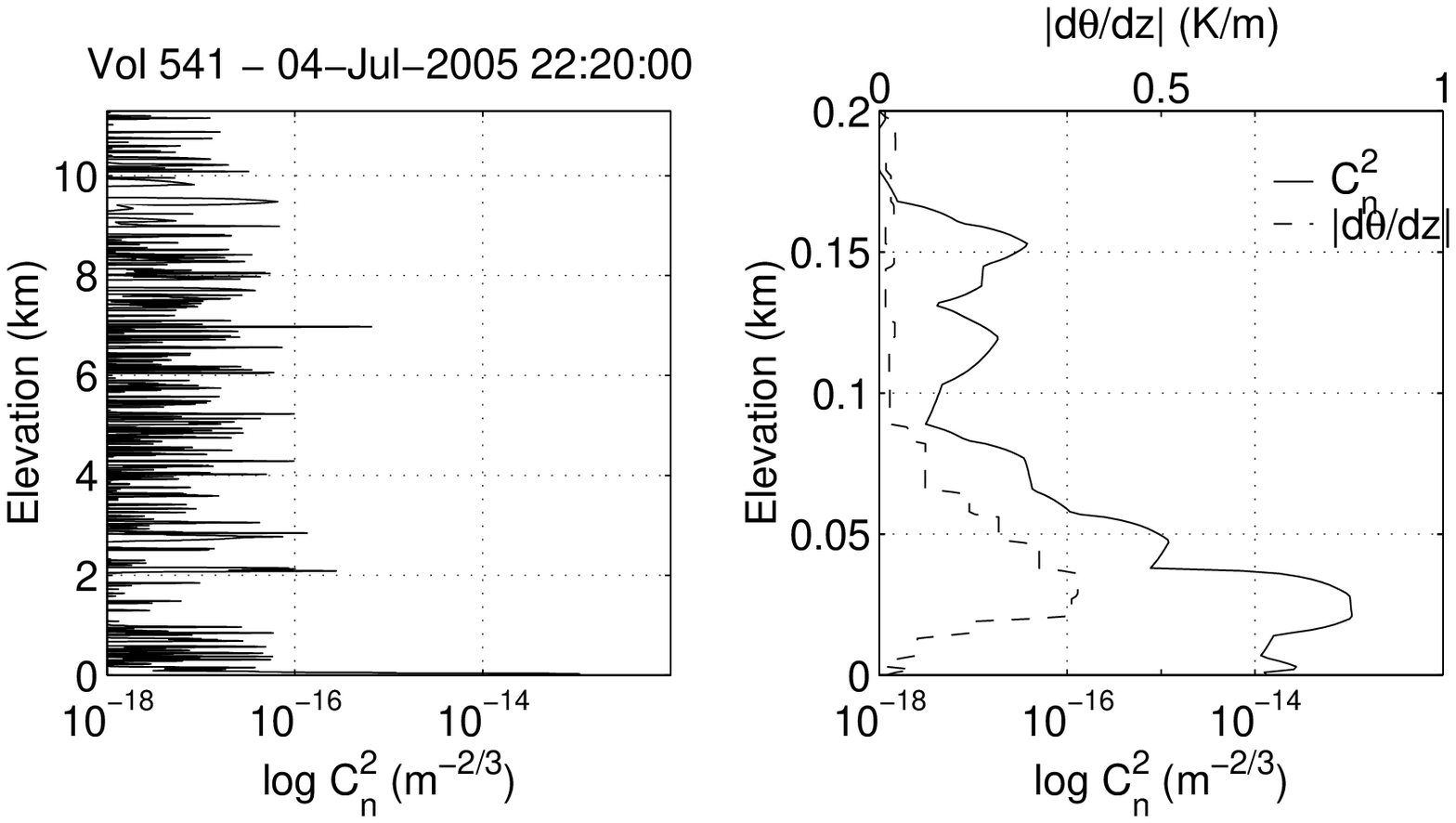}\\
\includegraphics[width=\figwidth]{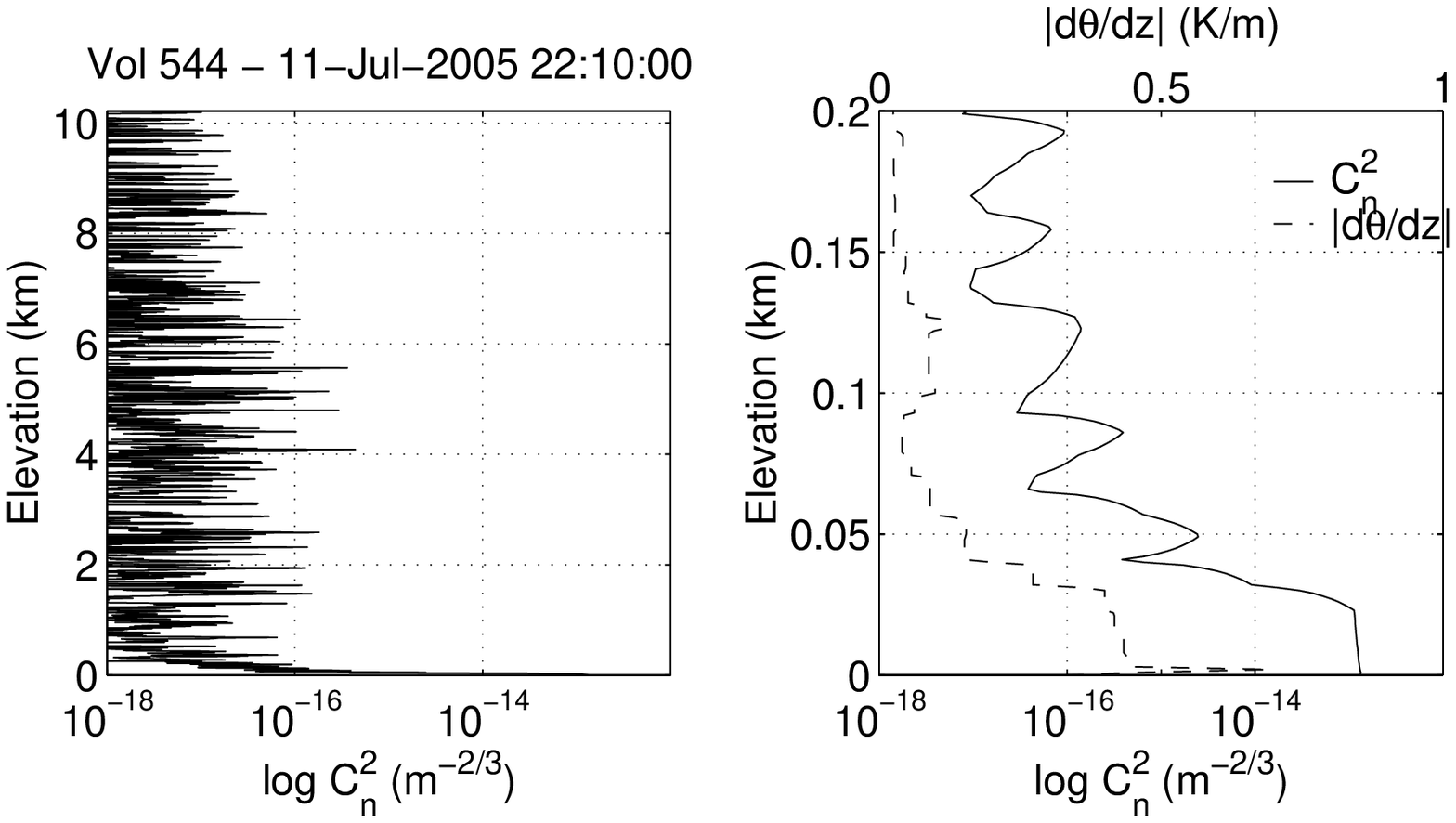}\ \hskip 1cm 
\includegraphics[width=\figwidth]{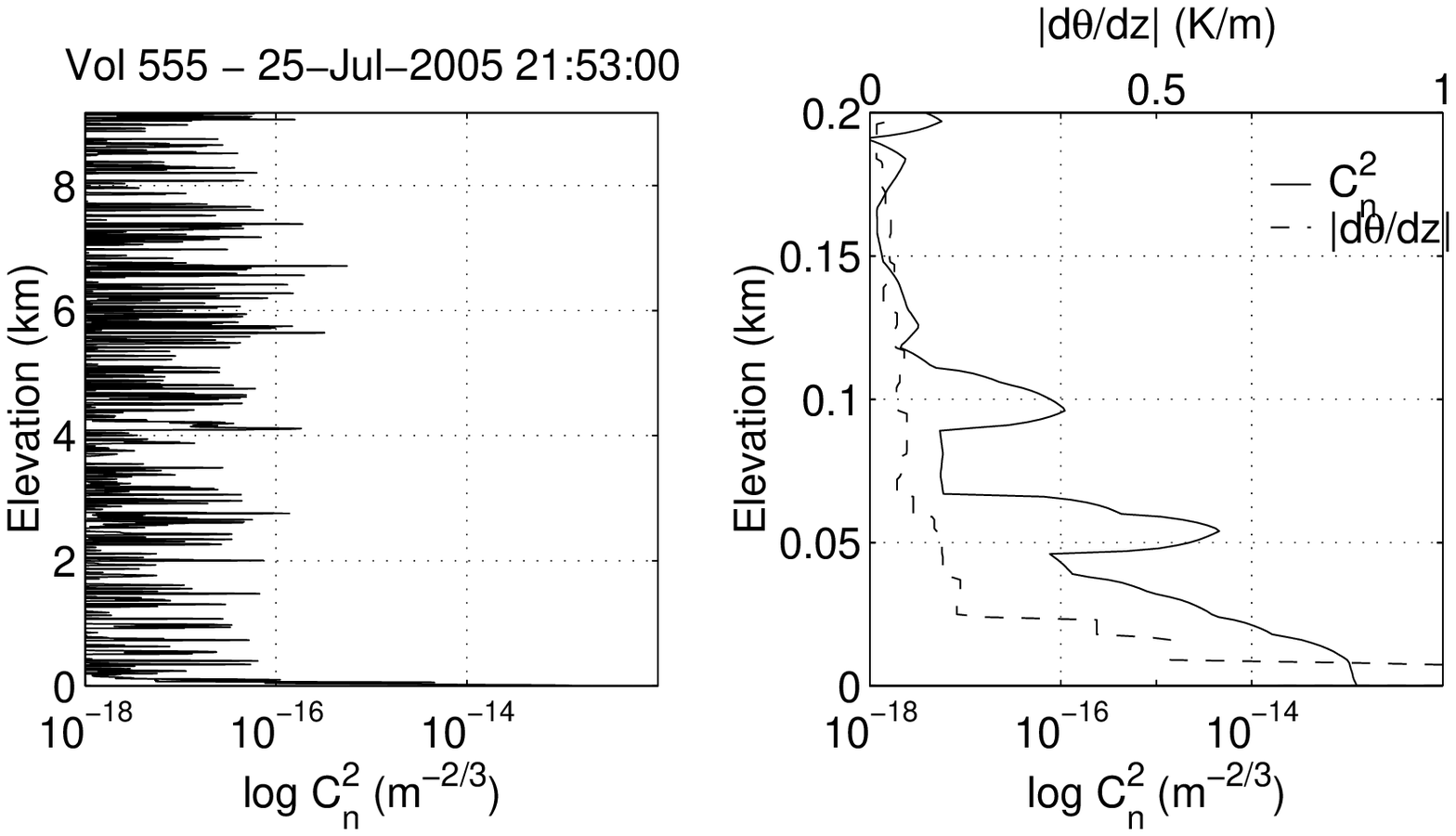}\\
\caption{Individual characteristic profiles of $C_n^2$ and of the potential temperature gradient $d\theta/dz$ for 4 balloons.}
\label{fig:cn2indiv}
\end{figure*}

\begin{figure}
\includegraphics[width=\figwidth]{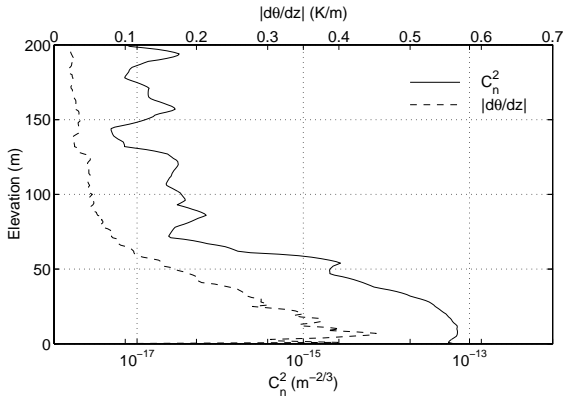}
\caption{Average vertical profiles of $C_n^2$ and of the potential temperature gradient $d\theta/dz$ in the first 200~m above ground.}
\label{fig:cn2gratemp}
\end{figure}

We started to launch balloons on March, 15$^{\rm th}$ 10:25pm when the Sun elevation was below -12\ddeg. All further launches were performed at night between 10 and 11pm local time. At the date of writing this paper, 16 balloons have been succesfully launched and provided exploitable vertical profiles $C_n^2(h)$. Averaged profiles are shown on fig.~\ref{fig:cn2vsalt}b. The seeing at a given altitude has been computed from $C_n^2(h)$ profiles (fig.~\ref{fig:cn2vsalt}c). Individual plots of 4 typical profiles are shown in fig.~\ref{fig:cn2indiv}. Largest values around $10^{-13}$~m$^{-2/3}$ are found just above the ground. The remaining of the turbulent energy being well distributed with the altitude, with values around $10^{-18} m^{-2/3}$ at the highest elevations ($h=15$~km). Average $C_n^2$ profile in the first 200~m is plotted together with the potential temperature gradient $d\theta/dz$ (figure~\ref{fig:cn2gratemp}). This gradient appears in the definition of the Richardson number (eq.~5 of Marks \etal , 1999) that describes the apparition of turbulence. The similarity between the two curves is remarkable. Wind speed gradient, also appearing in the Richardson number, could not be estimated because of lack of data near the ground.

\begin{figure}
\includegraphics[width=\figwidth]{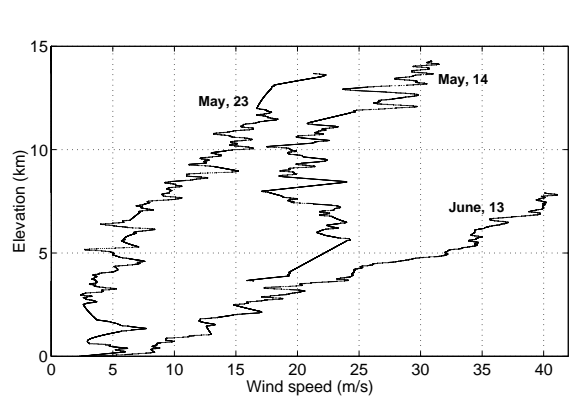}
\caption{Wind speed profiles measured by 3 radiosoundings in May and June.}
\label{fig:windspeed}
\end{figure}

We found a ground seeing, above ground and up to 15--20~km, of 1.9$\pm$0.5 arcsec, and a 36$\pm$10~m  thick surface layer accounting for 87\% of the turbulent energy (the integral of $C_n^2$). The seeing above this surface layer is 0.36$\pm$0.19~arcsec. The upper limit of the surface layer is defined as in Marks \etal, 1999: it is the altitude at which successive calculations of the seeing differ from less than 0.001\arcsec. Other parameters deduced from individual profiles are summarized in table~\ref{table:optparam}. Isoplanatic angle $\theta_0$ and coherence time $\tau_0$ correspond to the adaptive optics definition (eq.~7 and 9 of Marks \etal , 1999). All values compued above the surface layer are in remarkable agreement with measurements reported by Lawrence \etal\ (2004). $\theta_0$ appears to be smaller than the summer median value of 6.8~arcsec, though the surface layer is of little influence on it. High altitude strong winds (up to 30~m/s at $h$=16km) have indeed been observed in May and June, as illustrated by Fig.~\ref{fig:windspeed}. The correlation between these high wind speeds and the optical turbulence parameters is a study which will be adressed in a forthcoming paper when more data will be available.


\subsection{Seeing and isoplanatic angle monitorings}
\begin{figure}
\includegraphics[width=\figwidth]{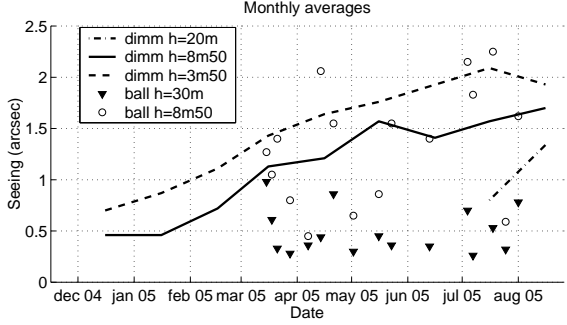}
\caption{Monthly averaged ground seeing from the monitors located at elevations $h=3.5$~m, $h=8.5$~m, and $h=20$~m. Individual points are ballon-based estimations at $h=8.5$~m and $h=30$~m.}
\label{fig:seeingmens}
\end{figure}

\begin{figure}
\includegraphics[width=\figwidth]{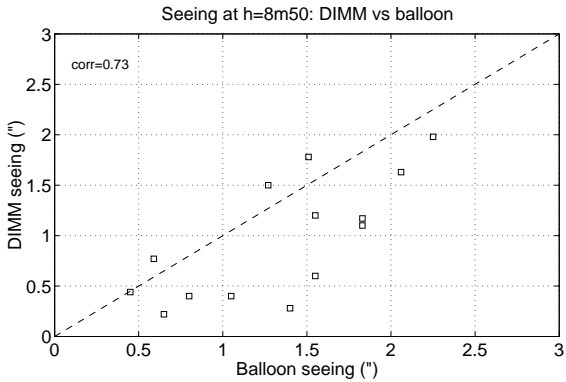}
\caption{Plot of the seeing estimated by the DIMM at $h$=8m50 versus seeing computed from the balloons profiles at the same height and around the same time, when available.}
\label{fig:seeingballvsdimm}
\end{figure}

The seeing data taken into account in this paper have been collected during the period March 1$^{\rm st}$ -- August 23$^{\rm th}$ , 2005. The seeing statistics provided in Table~\ref{table:optparam} stands for the monitor located on the platform (elevation $h=8.5$~m). The ``ground'' seeing monitor at $h=3.5$~m has also been running during the same period of time, while the ``roof'' monitor at $h=20$~m provides data since July, 23rd. Figure~\ref{fig:seeingmens} shows, for the two monitors, the monthly averaged seeing evolution during the transition from summer to winter. Both seeings follow the same positive trend. Figure~\ref{fig:seeingballvsdimm} shows a plot of the seeing integrated from $C_n^2$ profiles from $h=8.5$~m, as a function of the DIMM seeing at $h=8.5$~m, taken at the same time and averaged over the duration of the flight (typically two hours). The two data sets are consistent, with a correlation of 0.73. The dashed line on the graph is the first bissector.

Isoplanatic angle monitoring started later; data are available since May, 19$^{\rm th}$ and statistics are displayed on table~\ref{table:optparam}. Here again, data from the monitor are compatible with data from the ballon $C_n^2$ profiles. $\theta_0$ appears to be similar to South Pole value of 3.23~arcsec (Marks \etal , 1999).



\section{Conclusion}
\label{par:concl}
The situation at Dome C appears to be similar at South Pole (Marks \etal , 1999): a poor ground seeing mainly due to a strongly turbulent boundary layer. In both sites the free atmosphere seeing is around 0.35~arcsec, the isoplanatic angle around 3~arcsec. The main difference being the height of the boundary layer~: 220 m at South Pole and 36~m at Dome C. This boundary layer seems to be mainly due to the strong thermal gradient near the ground (around 20\ddeg over 100~m). This gradient is likely to exist on the whole Antarctic plateau and we might find a similar boundary layer everywhere. The question is the thickness of this layer. At Dome~C we could imagine to build 30~m high structures and put telescope on top to benefit from the excellent free atmosphere seeing. This is not an insurmountable problem: existing telescopes are often elevated (ESO 3.6m at La Silla is at 30~m, CFHT at Mauna Kea is at 28~m).

As Lawrence et al (2004) state in their paper, the MASS + SODAR profiling 
technique does not sense the first 30m of the atmosphere.  Our measurement 
now show that, unfortunately, this layer is very turbulent and, for a 
telescope mounted at ice level, contributes 87\% of the total seeing.  Thus, 
although the free atmosphere seeing at Dome C is excellent, it is important 
to appreciate that the surface layer (which we conclude is typically 36 m 
thick) results in very poor seeing for a telescope at ice level.

The properties of the turbulence in the boundary layer need to be investigated carefully. For this purpose, four micro-thermal sensors pairs were set up onto the 32~m-high American tower, to estimate and monitor the ground layer turbulence, at elevations  $h=$2m60, 8m10, 16m and 32m. Results from this experiment are expected soon, as well as other measurements during the rest of the winter.

\section*{Acknowledgements}
We wish to thanks the summer camp logistics and the winterover team at Concordia for their help. The program is supported by the Polar Institutes IPEV and ENEA.

\begin{table}

\begin{tabular}{l|c|c|c}
\multicolumn{4}{c}{\it Balloon data}\\ \hline
 & $\epsilon$ & $\theta_0$ & $\tau_0$ \\
Elevation         & (arcsec) & (arcsec)      &  (ms)   \\ \hline
$h\ge$8.5 m & 1.4$\pm$0.6   & 4.7$\pm$2.6 & 2.9$\pm$7.0 \\
$h\ge$30 m & 0.36$\pm$0.19	 & 4.7$\pm$2.6 & 8.6$\pm$7.1 \\ \hline
\multicolumn{4}{c}{ }\\
\multicolumn{4}{c}{\it Monitors data}\\ \hline
 & $\epsilon$ & \multicolumn{2}{c}  {$\theta_0$}  \\
  $h\ge$8.5 m & (arcsec) & \multicolumn{2}{c} {(arcsec) }    \\ \hline
Median & 1.3$\pm$0.8 & \multicolumn{2}{c} {2.7$\pm$1.6 }\\
Min/Max & 0.12/3.37 & \multicolumn{2}{c} {0.43/10.91}\\
Ndata & 36127 & \multicolumn{2}{c} {9501}\\ \hline

\end{tabular}

\caption{Median optical parameters at two elevations above ground. Values are computed from 16 individual $C_n^2$ profiles. The uncertainties are standard deviations of the values. }\label{table:optparam}

\end{table}

\end{document}